\documentclass{aa}
\usepackage{graphicx,color,subfigure,multirow}
\usepackage{txfonts}
\usepackage{verbatim}
\graphicspath{{./figures/}}
\usepackage{natbib,twoopt}
\usepackage[breaklinks=true]{hyperref} 

\hypersetup{
        colorlinks=true,   
        urlcolor=blue,     
        linkcolor=blue     
}
\bibpunct{(}{)}{;}{a}{}{,} 

\makeatletter
\newcommand{\bibnote}[2]{\global\@namedef{#1note}{#2}}
\newcommand{\biblink}[2]{\global\@namedef{#1link}{#2}}
\makeatother
\makeatletter
\protected\def\stonyslink{%
        \def\hyper@linkstart##1##2{}\let\hyper@linkend\@empty}
\newcommandtwoopt{\citeads}[3][][]{%
        \href{http://adsabs.harvard.edu/abs/#3}%
        {\stonyslink \citealp[#1][#2]{#3}}
        \biblink{#3}{\href{http://adsabs.harvard.edu/abs/#3}{ADS}}}
\newcommandtwoopt{\citepads}[3][][]{%
        \href{http://adsabs.harvard.edu/abs/#3}%
        {\stonyslink \citep[#1][#2]{#3}}
        \biblink{#3}{\href{http://adsabs.harvard.edu/abs/#3}{ADS}}}
\newcommandtwoopt{\citetads}[3][][]{%
        \href{http://adsabs.harvard.edu/abs/#3}%
        {\stonyslink \citet[#1][#2]{#3}}
        \biblink{#3}{\href{http://adsabs.harvard.edu/abs/#3}{ADS}}}
\newcommandtwoopt{\citeyearads}[3][][]{%
        \href{http://adsabs.harvard.edu/abs/#3}%
        {\stonyslink \citeyear[#1][#2]{#3}}
        \biblink{#3}{\href{http://adsabs.harvard.edu/abs/#3}{ADS}}}
\makeatother

\begin{document}

\title{Lunar ejecta as the missing piece to resolving lunar cratering asymmetry}
\author{
    Hailiang Li\inst{1}
        \and
    Xiaoping Zhang\inst{1}
    \and
        Li-Yong Zhou\inst{2}\fnmsep\inst{3}
        }
\authorrunning{Li et al.}
\institute{State Key Laboratory of Lunar and Planetary Sciences, Macau University of Science and Technology, Macau 999078, China\\
            \href{mailto:xpzhang@must.edu.mo}{xpzhang@must.edu.mo}
            \and
            School of Astronomy and Space Science, Nanjing University, 163 Xianlin Avenue, Nanjing 210046, China
              \and
            Key Laboratory of Modern Astronomy and Astrophysics in Ministry of Education, Nanjing University, China
        }
\date{}

\abstract{The leading-trailing asymmetry in lunar crater distribution provides a critical record of inner Solar System dynamics, yet the long-standing discrepancy between the observed higher asymmetry and lower theoretical predictions indicates a gap in our understanding of the impactor population. This paper hypothesizes that lunar impact ejecta, which can enter Earth-like orbits and later exit, constitute a  component that had previously been unaccounted for. Using numerical simulations, we found that $\sim$25\% of escaped ejecta will re-impact the Earth-Moon system within 3 Myr, with about 1.2\% striking the Moon. Crucially, these lunar impacts exhibit an extreme leading-trailing asymmetry, with a ratio of 5.9. Our results indicate that the assumption of lunar ejecta, comprising $\sim$15\% of total impactors can indeed explain the observed asymmetry, leading to their recognition as active agents shaping the lunar impact record. This work provides new constraints on our understanding of the impact environment of the Earth-Moon system, with direct relevance to the interpretation of lunar geology, the transport of lunar material to Earth, and ongoing space exploration missions.
}

\maketitle{}

\section{Introduction}

The lunar surface, largely untouched by geological recycling processes, serves as a pristine archive of the inner Solar System's impact history. The distribution of craters across this surface is not random; it encodes fundamental information about the dynamics and populations of small bodies that have interacted with the Earth-Moon system over billions of years \citepads[e.g.,]{Halliday1964, Wood1973, Halliday1982, Pinet1985, Werner2010}. A key feature of this distribution is the leading-trailing asymmetry, a direct consequence of the Moon's tidal locking.

Due to synchronous rotation, the Moon's leading hemisphere constantly faces its direction of orbital motion, acting as a ``windshield'' that intercepts more impactors than the trailing hemisphere. Furthermore, higher impact velocities on the leading side generate larger craters. Together, these factors naturally result in a higher crater density on the leading side \citepads[e.g.,][]{LeFeuvre2011}. This phenomenon was first observed and studied in other satellites \citepads[e.g.,][]{Zahnle1998, Schenk1999, Zahnle2001} before being confirmed for the Moon \citepads{Morota2003}. In this paper, we quantify this asymmetry using the crater (impact) ratio between the leading and trailing sides (denoted as L/T). 

Previous observational studies have established several estimates for the L/T. Based on images of the lunar surface captured by the Clementine spacecraft, \citetads{Morota2003} statistically analyzed radial craters larger than 5\,km diameter, obtaining an L/T of 1.44. From the lunar nighttime temperature maps obtained from thermal infrared observations, \citetads{Williams2018}  found a close association between temperature anomalies (i.e., cold spots) and craters. By quantifying the number of cold spots, an L/T of about 1.4 was derived. Furthermore, crater counting on lunar highlands shows that the asymmetry of lunar craters had formed as early as 4 Gyr ago \citepads{Zhao2024}. Notably, a significant L/T of 3.11 during the Imbrian Nectarian (approximately 3.85 to 3.8 Gyr ago) has been observed.

A critical issue relates to the size of craters. \citetads{Williams2018} have indicated that smaller cold spots exhibit greater asymmetry, suggesting that smaller impactors may possess distinct orbital distributions. Furthermore, the impact events currently recorded on the lunar surface  predominantly come from smaller scale meteoroids and micro-impacts. The micro-impact data detected during the Apollo passive seismic experiments from 1969 to 1977 reported L/T values of 1.4 to 1.9 \citepads{Kawamura2011}, while impact flash data have indicated an L/T ratio of about 1.5 \citepads{Oberst2012}.

Although the asymmetry has been confirmed by multiple previous studies, involving diverse observational techniques and examined craters or impacts of varying ages and sizes, their results are largely consistent; with L/T values ranging from approximately 1.4 to 1.9 from various studies. However, a significant discrepancy persists between observations and theoretical models. These models, which primarily consider near-Earth objects (NEOs) as impactors, predict a much milder asymmetry (L/T$\sim$1.14; e.g., \citeads[][]{Gallant2009, Ito2010, LeFeuvre2011, Wang2016}). This gap between observation and theory points to a critical missing component in our understanding of the impactor flux: a population of low-velocity impactors on Earth-like orbits.

The degree of leading-trailing asymmetry is highly sensitive to the impactor's encounter velocity ($v_{enc}$) relative to the Moon's orbital velocity ($v_{M}$), with lower $v_{enc}$ leading to a more pronounced asymmetry \citepads[e.g.,][]{Horedt1984, Zahnle2001, LeFeuvre2011, Wang2016, LiHuacheng2021}. The observed L/T values suggest the impactors with $v_{enc}$ about 8–14\,km/s. However, NEOs typically approach the Earth-Moon system with high velocities ($v_{enc}\sim$20\,km/s), which dilutes the asymmetry effect. To reconcile observational constraints, the presence of numerous low-velocity impactors (often referred to as low-velocity NEOs) is required. These objects are characterized by their Earth-like orbits, yet due to the complex dynamical environment of near-Earth space, their survival timescales are typically short \citepads[e.g.,][]{Michel1997, Fenucci2023}, necessitating specific and efficient replenishment mechanisms. Several hypothetical candidates of supplements have been put forward, for example, the tidal disruption debris of small bodies \citepads{Ito2010} and the Earth's co-orbital objects \citepads{Malhotra2019}.

Another compelling and natural candidate for these low-velocity impactors is lunar impact ejecta. Material ejected from the Moon during impacts naturally possesses orbits similar to Earth's, with low relative velocities upon return. The recent hypothesis that the Earth quasi-satellite Kamo'oalewa could be lunar ejecta from the Giordano Bruno crater has significantly renewed interest in the dynamical pathways of lunar material \citepads{Sharkey2021, Jiao2024}. While previous studies have explored the orbital evolution of lunar ejecta \citepads[][]{Gladman1995, CastroCisneros2023, CastroCisneros2025a, CastroCisneros2025b, Sfair2025}, a comprehensive, quantitative assessment of their contribution to the lunar cratering record (particularly the leading-trailing asymmetry) has been lacking.

This paper aims to conduct high-resolution numerical simulations to track the dynamical evolution of lunar ejecta over million-year timescales and to provide further understanding of the asymmetric distribution of lunar craters. We introduce our simulation settings and methodology in Sect. 2. In Sect. 3, we  show the unique orbital characteristics of lunar ejecta and their correlation with initial conditions, analyze their long-term orbital excitation, calculate the flux of such material returning to the Earth-Moon system through re-impact, and examine the L/T from such events. Section 4 provides discussion and conclusions.

\section{Methods}

\subsection{Initial conditions}
In our simulation, the Solar System model includes the Sun, eight planets, and the Moon. The initial positions of all major bodies are based on osculating elements at epoch MJD\,60000.0. At this moment, the Moon is located behind Earth along its orbital velocity direction, with the Sun-Earth-Moon angle approximately $62.5^{\circ}$.

A critical issue concerns the position, velocity, and angle at which ejecta launch from the Moon, which involves dynamics associated with impacts \citepads{Housen2011}. Previous numerical simulation works have chosen a few representative launch positions \citepads[e.g.,][]{CastroCisneros2023, Jiao2024, CastroCisneros2025a, CastroCisneros2025b}. In this study, we aim to test small bodies launched from various locations to gain a comprehensive understanding of the evolution of lunar ejecta. Therefore, we employed a Fibonacci spherical sampling method to uniformly sample the entire lunar surface.

Typically, ejecta form a conical launch pattern \citepads[e.g.,][]{Gladman1995, Jiao2024, CastroCisneros2025b}, with an angle of approximately $45^\circ$ between the motion direction of the ejecta and the lunar surface normal vector.  \citetads{Sfair2025} investigated cases where ejecta were launched outward along the normal vector and found no significant difference in statistical outcomes, compared to randomly selected launch angles. Consequently, as the precise launch angle falls outside the scope of our investigation, we accordingly implemented a model where all particles are launched perpendicular to the lunar surface, initiating their trajectories at a distance of two lunar radii from the center.

The escape velocity on the lunar surface is approximately 2.38\,km/s, while at two lunar radii, the minimum velocity required to escape lunar gravity is $\sim$1.68\,km/s. According to \citetads{Jiao2024}, the number of ejecta follows a power law of exponent -4 with the launch velocity ($v_L$), the vast majority of ejecta have $v_L$ values falling within the range of 2.38\,km/s to 6\,km/s. In this study, we denote the initial velocity of the small bodies as $v_0$, considering six different $v_0$ values ranging from 2\,km/s to 6\,km/s. We note that our adopted $v_0$ represents the velocity at twice the lunar radius. Taking into account the deceleration during the trajectory from the lunar surface to this altitude, the corresponding $v_L$ values for the six simulations range from 2.61\,km/s to 6.23\,km/s, which is highly representative of the $v_L$ of lunar ejecta. Each simulation group comprises 10,000 small bodies, with appropriate weighting applied in subsequent calculations of collective effects.

\subsection{Orbital dissimilarity metric}

The relative velocity at which small bodies enter the Earth-Moon system is a crucial factor influencing the leading-trailing asymmetry in lunar crater distribution. Low-velocity NEOs typically occupy the Earth-like orbits. It is reasonable to assume that if a small body originates as lunar ejecta, it should possess an orbit similar to that of the Earth. We go on to  define a metric to quantify how closely the orbit of a small body resembles the Earth's orbit. Assuming the Earth's orbit is circular with a semimajor axis of 1\,au, for any small body with known $a$, $e$, and $i$, appropriate values for $\Omega$, $\omega$, and $M$ can be selected to achieve an Earth encounter. The relative velocity, $\Delta v$, during this encounter serves as our measure of orbital similarity. Since a higher $\Delta v$ implies a lower degree of orbital similarity, we refer to this metric as orbital dissimilarity metric.

\begin{figure}[!htb]
\centering
\resizebox{\hsize}{!}{\includegraphics{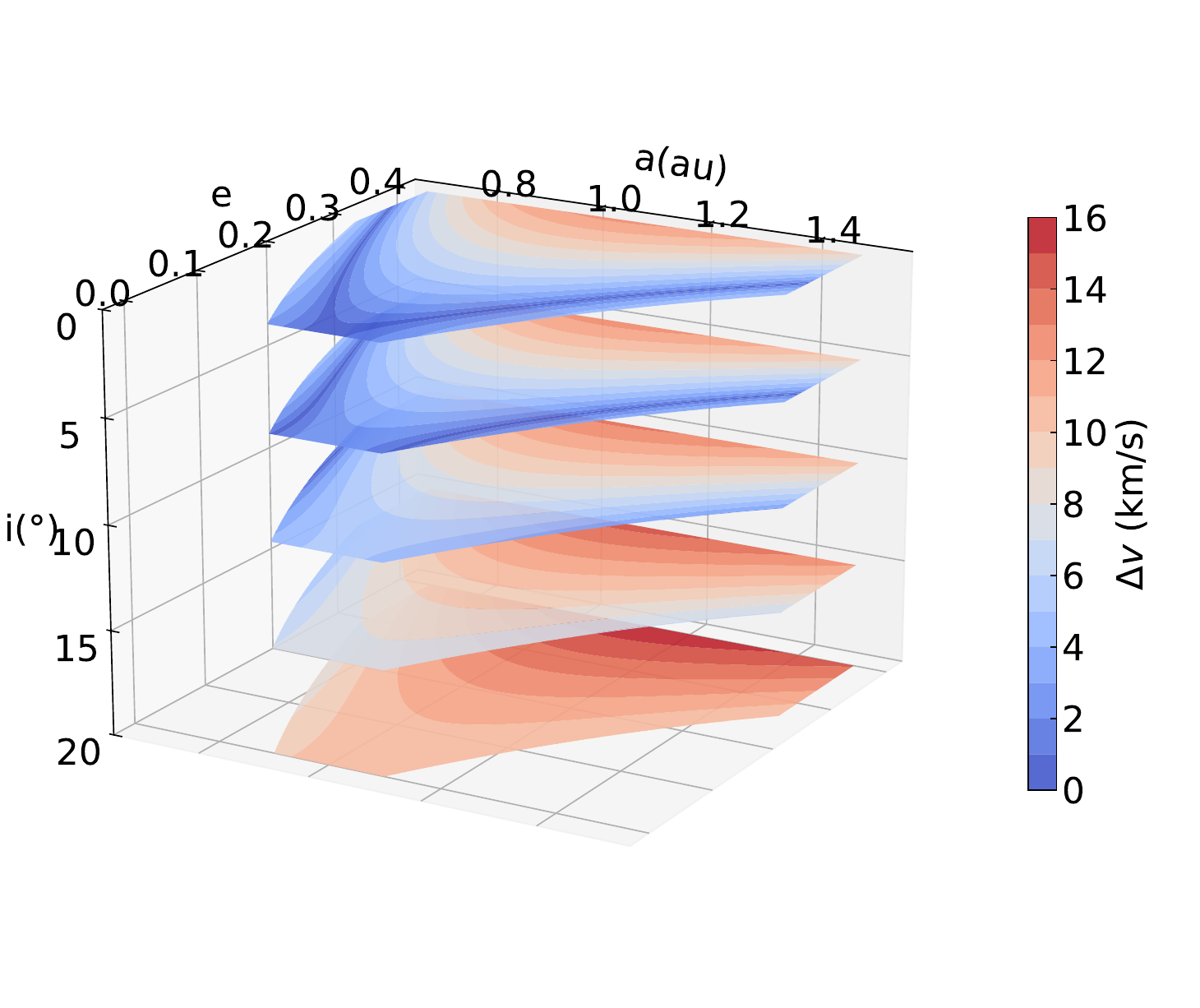}}
\caption{Orbital dissimilarity metric across the $a-e-i$ parameter space. Bluer colors indicate higher degrees of orbital similarity.}
\label{fig:dv}
\end{figure}

To calculate the $\Delta v$, we leverage rotational symmetry by assuming the encounter occurs when Earth transits the vernal equinox. To position the small body at this point as well, we can (without any loss of generality) set $\Omega=0$, the true anomaly $f=-\omega$, and the heliocentric distance $r=1$\,au. If we denote Earth's velocity as $v_E$, and the velocity components of the small body as $\dot{x}$, $\dot{y}$, and $\dot{z}$, then the $\Delta v$ can be expressed as $\sqrt{\dot{x}^2+(\dot{y}-v_E)^2+\dot{z}^2}$. Expressing $\dot{x}$, $\dot{y}$, and $\dot{z}$ in terms of $a$, $e$, and $i$, we can derive the following equations,
\begin{equation}
    \Delta v=\sqrt{3-\frac{1}{a}-2\sqrt{a(1-e^2)}\cos(i)}=\sqrt{3-T_E}
    \label{eq:dv},
\end{equation}
where $T_E$ represents the Tisserand parameter of the small body with respect to Earth and $\Delta v$ is in units of Earth's orbital velocity ($\sim$29.8\,km/s). Equation \ref{eq:dv} is meaningful only when $T_E<3$ and our derivation requires that the orbit of the small body satisfy that $a(1-e)<1$\,au and $a(1+e)>1$\,au. In practical calculations, we slightly extended the applicability of Eq. \ref{eq:dv}. For all orbits satisfying $a(1-e)<1.1$\,au and $a(1+e)>0.9$\,au, we used $\Delta v=\sqrt{|3-T_E|}$ to compute their orbital dissimilarity metric. Figure~\ref{fig:dv} illustrates the $\Delta v$ between the small body and Earth for different $a$, $e$, and $i$ values.

Figure~\ref{fig:dv} shows that the highest degree of orbital similarity occurs when the orbit of the small body perfectly matches that of Earth (i.e., $a=1$\,au and $e=i=0$). Generally, for a given $a$, a larger $e$ leads to a greater $\Delta v$. However, with a fixed $e$, the $\Delta v$ does not always reach its minimum precisely at $a=1$\,au, because the encounter geometry between the two objects becomes less favorable; a marginally larger or smaller semimajor axis can improve orbital similarity by optimizing the encounter angle. On the other hand, the $\Delta v$ consistently increases with higher $i$.

\subsection{Calculation of L/T}

Regarding the distribution of lunar crater density, many previous studies have provided theoretical expressions \citepads[e.g.,][]{Zahnle2001, LeFeuvre2011, Wang2016}. While there are slight quantitative differences in the formulas presented in these works, they are   qualitatively consistent overall. The central point of the leading hemisphere of the Moon $(90^{\circ}W,0^{\circ})$ is commonly referred to as the apex, while the $(90^{\circ}E,0^{\circ})$ is termed the antapex. According to \citetads{Wang2016}, when considering only the leading-trailing asymmetry, the crater density $Nc$ at a given lunar surface location can be approximated as
\begin{equation}
    N_c=A_0(1+A_1 \cos(\beta))
    \label{eq:asy}
,\end{equation}
where $A_0$ represents the average crater density over the entire lunar surface, $A_1$ quantifies the degree of leading-trailing asymmetry, and $\beta$ denotes the spherical distance of the point from the apex. According to Eq. \ref{eq:asy}, the crater density ratio between the apex and antapex is $(1+A_1)/(1-A_1)$. Integrating Eq. \ref{eq:asy} over the entire lunar surface further yields the L/T as $(2+A_1)/(2-A_1)$. Evidently, the latter ratio is slightly smaller than the former. Both ratios have been employed in previous studies; in this work, we utilize the latter ratio as an indicator. For studies reporting apex-antapex density ratios, we converted them to the L/T to enable consistent comparisons.

In our simulations, the probability of small bodies impacting the Moon is significantly lower than the probability  of them impacting Earth. To enhance the statistical significance of the results, we did not directly track the positions of lunar impacts; instead, we considered all impacts on both the Earth and the Moon as ``potential lunar impactors.'' We recorded the relative velocity of two objects at the time of impact ($v_{imp}$) for each collision and used this information to deduce the $v_{enc}$ of the small body. For small bodies that collide with the Earth, the expression is $v_{enc} = \sqrt{v_{imp}^2 - \frac{2GM_E}{R_E} + \frac{2GM_E}{R_{H}}}$, where $G$, $M_E$, $R_E$, and $R_{H}$ represent the gravitational constant, Earth's mass, Earth's radius, and Earth's Hill radius, respectively. In this scenario, the influence of lunar gravity is neglected. On the other hand, for small bodies colliding with the Moon, the acceleration effect of Earth's gravity is taken into account. In this case, the expression for the encounter velocity is $v_{enc} = \sqrt{v_{imp}^2 - \frac{2GM_M}{R_M} - \frac{2GM_E}{a_M} + \frac{2GM_E}{R_{H}}}$, where $M_M$, $R_M$, and $a_M$ represent the Moon's mass, Moon's radius, and the semimajor axis of the Moon's orbit around the Earth, respectively.

After calculating the $v_{enc}$ of a small body, we can further derive the expected values of impacts on the leading and trailing sides of the Moon. \citetads{Wang2016} have developed a theoretical model for lunar crater distribution under a planar approximation (neglecting Earth and Moon gravity), from which we can approximate $A_1 \propto \frac{v_M}{v_{enc}}$. If focusing solely on crater number density, the proportionality constant is $\pi/2$; when accounting for the higher relative impact velocities (and thus larger crater sizes) on the leading side, the expression becomes
\begin{equation}
    A_1=\frac{\pi^2+(16-\pi^2)\gamma_p\alpha_p}{2\pi}\frac{v_M}{v_{enc}}
    \label{eq:A1}
,\end{equation}
where $\gamma_p$ and $\alpha_p$ represent parameters associated with crater formation and size distribution. With typical values, the coefficient in Eq. \ref{eq:A1} is approximately 2.53. Therefore, in this study, we adopted $A_1 = 2.53 \frac{v_M}{v_{enc}}$  to calculate the L/T.

Equation \ref{eq:A1} is utilized to calculate the expected probability of a small body impacting the leading and trailing sides. When computing the overall L/T for numerous objects, we count the expected impacts for both sides separately and then take their ratio. In cases where a small body's velocity is extremely slow such that $A_1 > 2$, we no longer apply the formula; instead, we assume the body will inevitably impact the leading side. Additionally, there are rare instances where the calculation of $v_{enc}$ yields a negative value under the square root. This typically occurs when the small body has experienced significant deceleration due to Earth's or Moon's gravity prior to impact. For such cases, we likewise consider the impact to be certain on the leading side.

\section{Results}

\subsection{Distinct orbits of escaped lunar ejecta}

\begin{figure}[!htb]
\centering
\resizebox{\hsize}{!}{\includegraphics{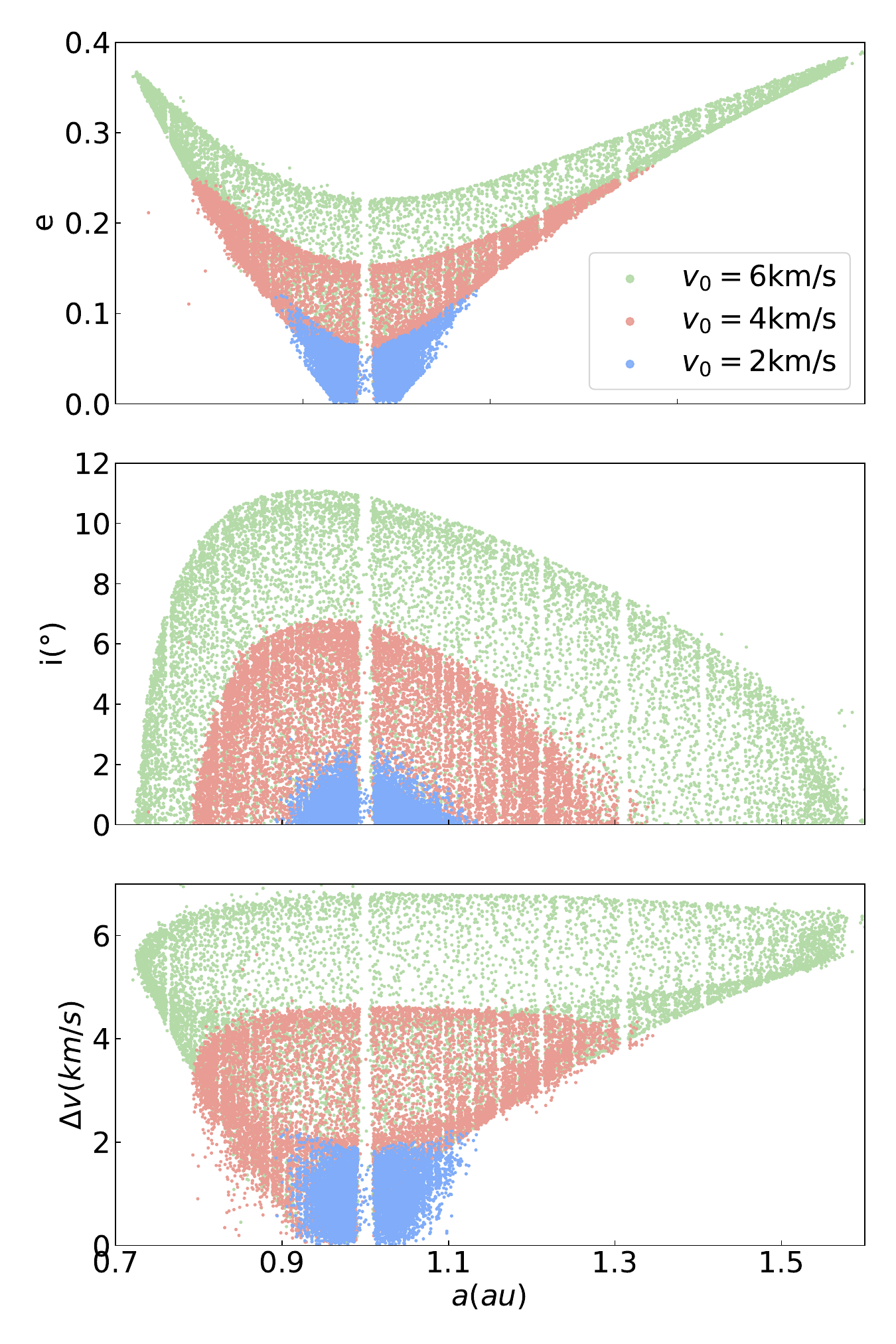}}
\caption{Distributions of $a$, $e$, $i$, and $\Delta v$  of simulated ejecta at $t = 100\,yr$. Points of different colors represent ejecta with different $v_0$ values.}
\label{fig:ini}
\end{figure}

Generally speaking, a significant portion of lunar ejecta will directly fall back onto the lunar surface and create craters. Due to the relatively slow impact velocities, these secondary craters typically have lower depth-to-diameter ratios and smaller sizes; thus, most of them have not been counted in previous statistics \citepads[e.g.,][]{Morota2003, Kawamura2011, Williams2018, Zhao2024}. In this study, we focus primarily on ejecta that can escape the gravitational influence of the Earth-Moon system. Consequently, our analysis considers only collisions and distributions of ejecta 100\,yr after the initial moment when they started from 2 radii distance. This timescale allows for the clearance of non-escaping material, while remaining short enough that most ejecta in heliocentric orbits have not yet experienced significant perturbations or re-impacts. Figure~\ref{fig:ini} presents the distribution of the $a$, $e$, $i$, and $\Delta v$ for the simulated ejecta population at $t = 100$\,yr.

As shown in Fig.~\ref{fig:ini}, ejecta with lower $v_0$ remain closer to Earth-like orbits (smaller $e$ and $i$, $a\sim1\,au$), resulting in a lower $\Delta v$. In contrast, ejecta with higher $v_0$ populate a broader range of orbital elements, achieving higher $\Delta v$ values. The maximum $\Delta v$ attainable is approximately equal to their $v_0$, demonstrating that initial conditions imprint a lasting kinematic memory. 

Moreover, when examining the three-dimensional (3D) trajectories of the ejecta, we notice that these trajectories almost converge at a single point in space. This point corresponds to the initial position of the Earth-Moon system in the simulation, which is where the ejecta were released. This configuration implies that whenever Earth approaches this point, the impact flux on both Earth and Moon would increase significantly. As this phenomenon repeats annually, it bears resemblance to meteor showers. Because of the orbital precession, the ejecta gradually disperse throughout the vast space near the Earth. Consequently, this orbital focusing effect can only persist for a limited duration, typically on the order of $10^3$ to $10^4$\,yr.

In Fig.~\ref{fig:ini}, distinct vertical gaps can be observed, which are induced by the Earth's mean-motion resonances (MMRs). The most prominent MMRs include: the 1:1 MMR at 1\,au, the 2:3 MMR at 1.31\,au, and the 3:4 MMR at 1.21\,au, along with numerous higher order MMRs that are also visibly present. This indicates that Earth's gravity constitutes the dominant mechanism influencing lunar ejecta, with significant effects already manifesting within a mere 100\,yr timescale.

\begin{figure*}[!htb]
\centering
\resizebox{\hsize}{!}{\includegraphics{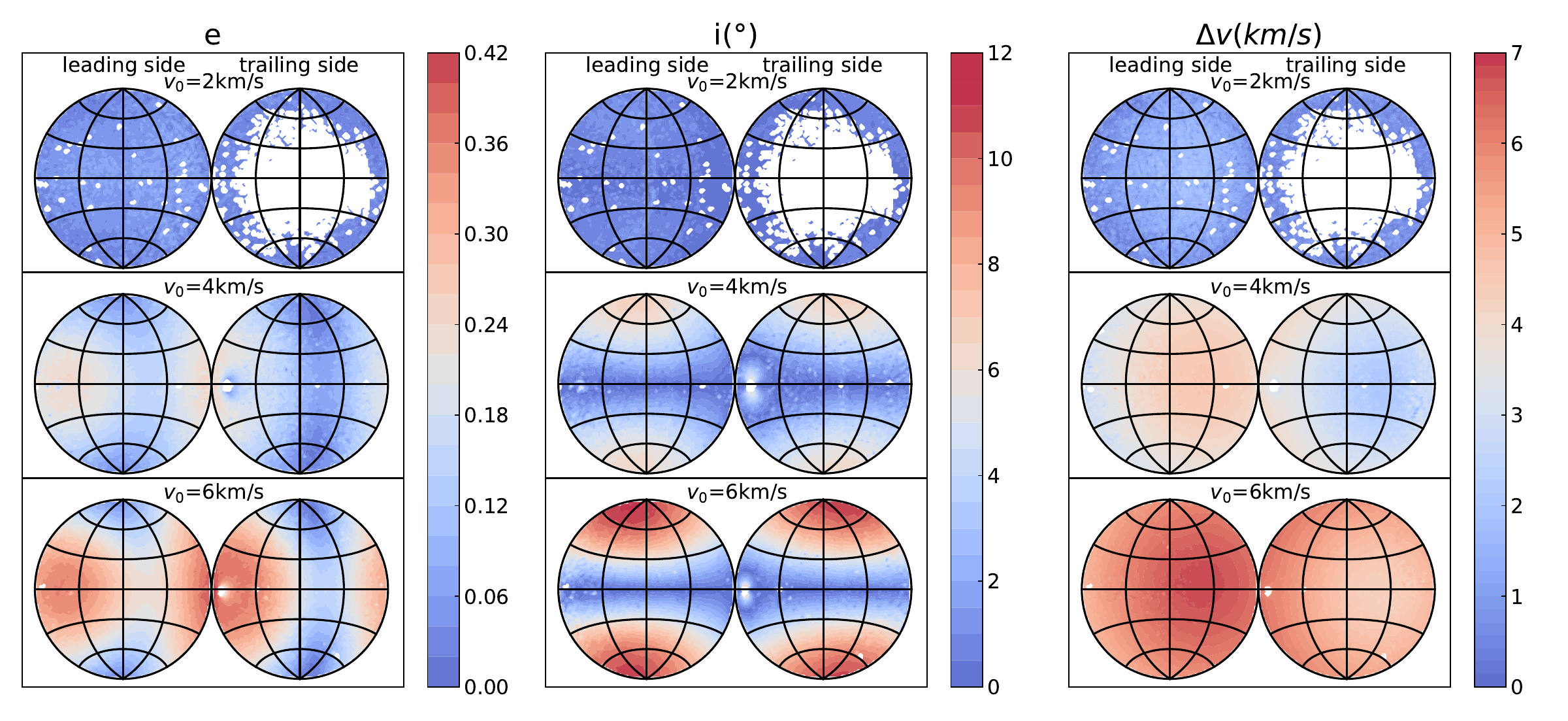}}
\caption{Relationship between the orbits of lunar ejecta and their initial positions at different $v_0$. From left to right, the panels represent $e$, $i$, and $\Delta v$, respectively. Blank areas indicate ejecta that did not survive until $t = 100\,yr$ in the simulations. Black lines mark the lunar latitude and longitude grid, with longitude intervals of $45^{\circ}$  and latitude intervals of $30^{\circ}$ .}
\label{fig:lat_v}
\end{figure*}

Another intriguing question concerns how the initial location and ejection time of lunar ejecta influence their orbital characteristics, which we explore in the next part. First, in Fig.~\ref{fig:lat_v}, we present the relationship between the orbits of lunar ejecta and their initial positions under three different values of $v_0$.

Figure~\ref{fig:lat_v} indicates that at lower $v_0$, the orbital characteristics of ejecta show minimal variation regardless of release position, as they all occupy Earth-like orbits with low-$e$ and low-$i$. However, for faster $v_0$, the release position exerts progressively greater influence on the final orbital parameters. In particular,  $i$ is largely determined by their latitude of ejection.

When $v_0$ = 2 km/s, it is evident that many small bodies in the trailing side of the Moon do not survive to 100\,yr. This phenomenon occurs because $v_0$ is defined relative to the Moon, while the ejecta from the leading side has a higher relative velocity with respect to Earth, making it more likely to escape Earth's gravitational influence. In the right panel of Fig.~\ref{fig:lat_v}, the $\Delta v$ of the ejecta are primarily affected by their $v_0$. However, for the same $v_0$, an asymmetry is noted between the leading and trailing sides, with objects ejected from the leading side generally exhibiting higher $\Delta v$.

For the cases where $v_0$ > 2 km/s, apart from some ejecta initially directed towards Earth (located near $(20^{\circ}E, 0^{\circ})$, with exact coordinates varying slightly with $v_0$), nearly all the ejecta can escape Earth's gravitational influence and survive for over 100\,yr. Furthermore, ejecta trajectories passing closer to Earth will be significantly affected by the scattering effect, manifesting as combinations of higher $i$ and lower $e$.

\begin{figure*}[!htb]
\centering
\resizebox{\hsize}{!}{\includegraphics{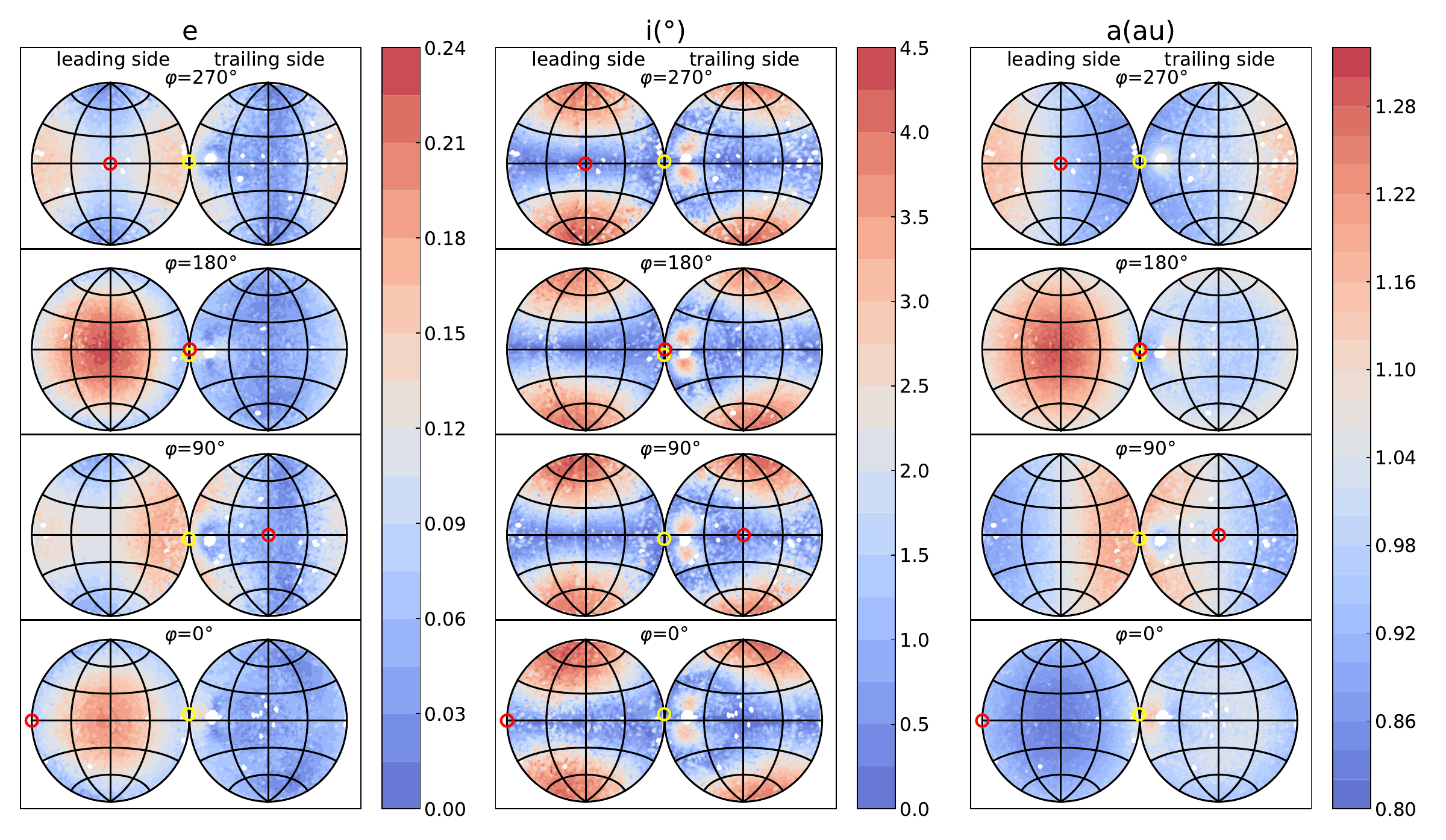}}
\caption{Similar to Fig.~\ref{fig:lat_v}, but the right column represents the semimajor axes of the ejecta and panels from top to bottom represent different lunar phases. In the plots, the $v_0$ of all lunar ejecta is 3 km/s. The open yellow and red circles mark the directions of Earth and the Sun, as viewed from the Moon, respectively.}
\label{fig:lat_v2}
\end{figure*}

On the other hand, the lunar phase at the time of ejecta launch also influences the simulation outcomes. To investigate this, we first defined the phase angle, $\phi$, as the Sun-Earth-Moon angle. We selected four time instances near the initial epoch of our primary simulation: MJD\,59988.7, MJD\,59995.3, MJD\,60002.3, and MJD\,60010.5. At these epochs, the $\phi$ approximates $270^{\circ}$, $0^{\circ}$, $90^{\circ}$, and $180^{\circ}$, respectively. We set the $v_0$ of the ejecta as 3\,km/s and examined the orbital differences of ejecta launched at these distinct times. The corresponding relationship is presented in Fig.~\ref{fig:lat_v2}. Obviously, the distribution of the inclination is basically not affected at all by the lunar phase.

In Fig.~\ref{fig:lat_v2}, the most apparent pattern is that the semimajor axes and eccentricities of small bodies are significantly influenced by the lunar phase. Ejecta launched at low latitudes and in directions perpendicular to the Sun (i.e., aligned with or opposite to Earth's orbital motion) tend to exhibit higher eccentricities. Among them, ejecta launched in the direction aligned with Earth's motion acquire both higher eccentricities and larger semimajor axes, while those launched opposite to Earth's motion are predominantly distributed on the inner side of Earth's orbit.

In addition to the velocity superposition effect caused by Earth's orbital motion, the Moon's revolution around Earth also leads to systematically higher eccentricities for ejecta released from the leading side. When the Moon moves in the same direction as Earth's orbital motion ($\phi = 180^{\circ}$), the two mechanisms reinforce each other. Ejecta from the leading side exhibit the highest eccentricities and most ejecta are distributed outside Earth's orbit. Conversely, when the Moon moves opposite to Earth's direction ($\phi = 0^{\circ}$), the effects partially cancel out, resulting in a notable reduction of high-eccentricity ejecta and causing most projectiles to fall inside Earth's orbit. In general, the average $\Delta v$ at $\phi = 180^{\circ}$ is approximately 26\% higher than at $\phi = 0^{\circ}$. 

The above discussion on lunar phases indicates that although our main initial epoch (MJD\,60000.0) was arbitrarily selected, the $\phi = 62.5^{\circ}$ at this time implies that the directions of Earth's and the Moon's motions were neither fully aligned nor opposed. This configuration ensures that the orbital distribution of lunar ejecta launched at this epoch retains a certain degree of representativeness. Furthermore, while our tests were conducted only at $v_0$ = 3\,km/s, it is anticipated that the influence of lunar phase would be relatively diminished at higher $v_0$. Additionally, the probability of impact events on the Moon may also vary slightly with lunar phase, suggesting that the quantity of ejecta emitted could also differ across phases. This mechanism warrants further investigation in future work.

\subsection{Long-term orbital excitation}

\begin{figure}[!htb]
\centering
\resizebox{\hsize}{!}{\includegraphics{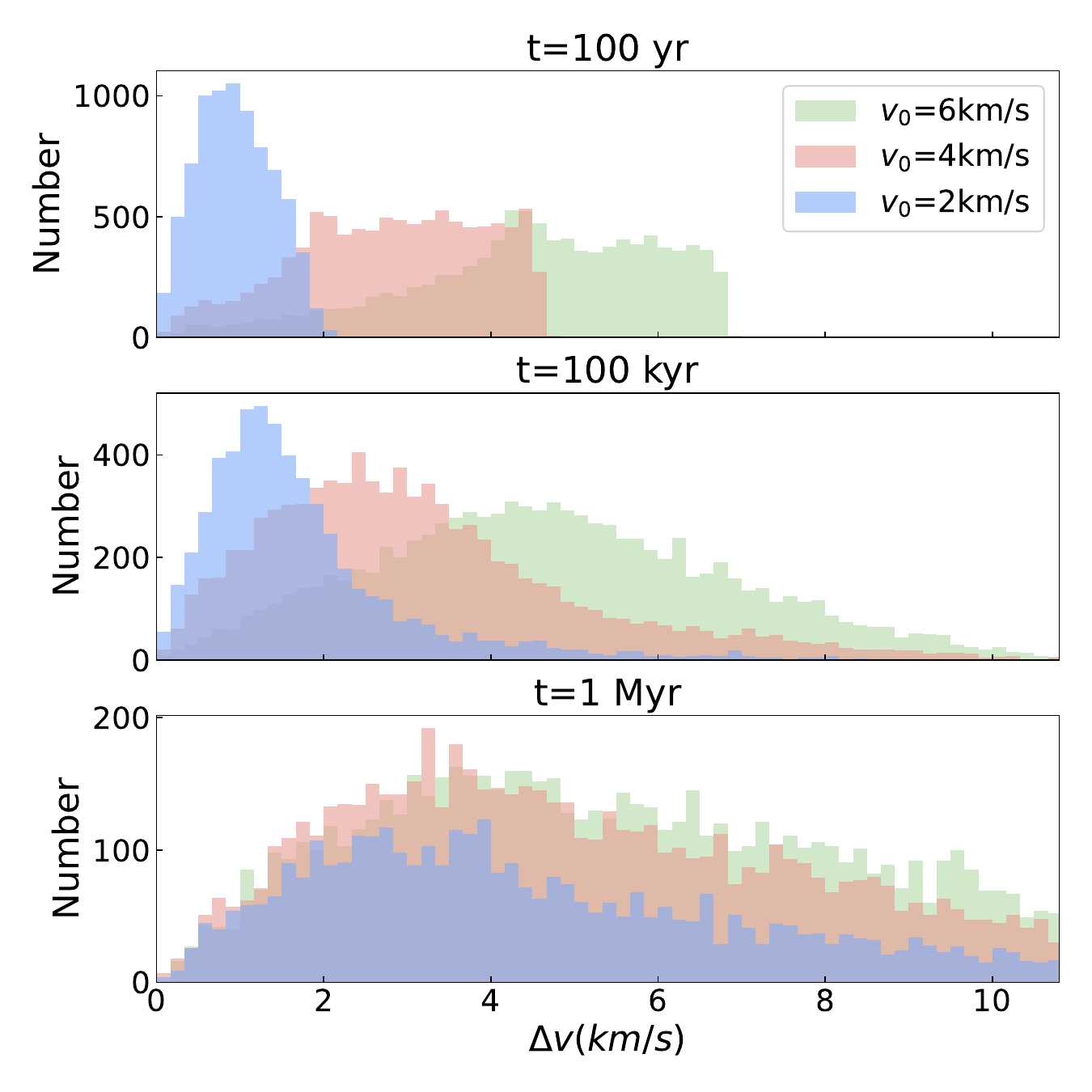}}
\caption{ $\Delta v$ distributions at 100\,yr, 100\,kyr, and 1\,Myr. Different colors representing distinct $v_0$. Only three values are plotted for clarity.}
\label{fig:v_dist}
\end{figure}

The subsequent evolution of lunar ejecta over million-year timescales shows progressive orbital excitation. In Fig.~\ref{fig:v_dist}, we present the $\Delta v$ distributions of lunar ejecta at three different times to illustrate their dynamical evolution.

The middle panel of Fig.~\ref{fig:v_dist} reveals that after 100\,kyr of evolution, although the quantity of surviving ejecta has decreased significantly, the average $\Delta v$ has increased only marginally, and differences among populations with different $v_0$ remain observable. However, the velocity distribution has undergone substantial changes, exhibiting a Rayleigh-like distribution pattern with an emergent population of higher $\Delta v$ objects. By 1\,Myr, the initially distinct $\Delta v$ distributions for different $v_0$ populations become largely indistinguishable, converging toward an average $\Delta v$ of $\sim$5\,km/s. However, it is noteworthy that these ejecta are still much slower than typical NEOs even after 1\,Myr.

This evolution follows a characteristic pattern: ejecta launched with low-velocities experience rapid depletion due to early collisions, while higher-velocity populations have longer lifetime but lower initial impact probabilities. The timescale for complete orbital randomization is approximately 1\,Myr, after which the dynamical memory of initial conditions becomes negligible.

\begin{figure*}[!htb]
\centering
\resizebox{\hsize}{!}{\includegraphics{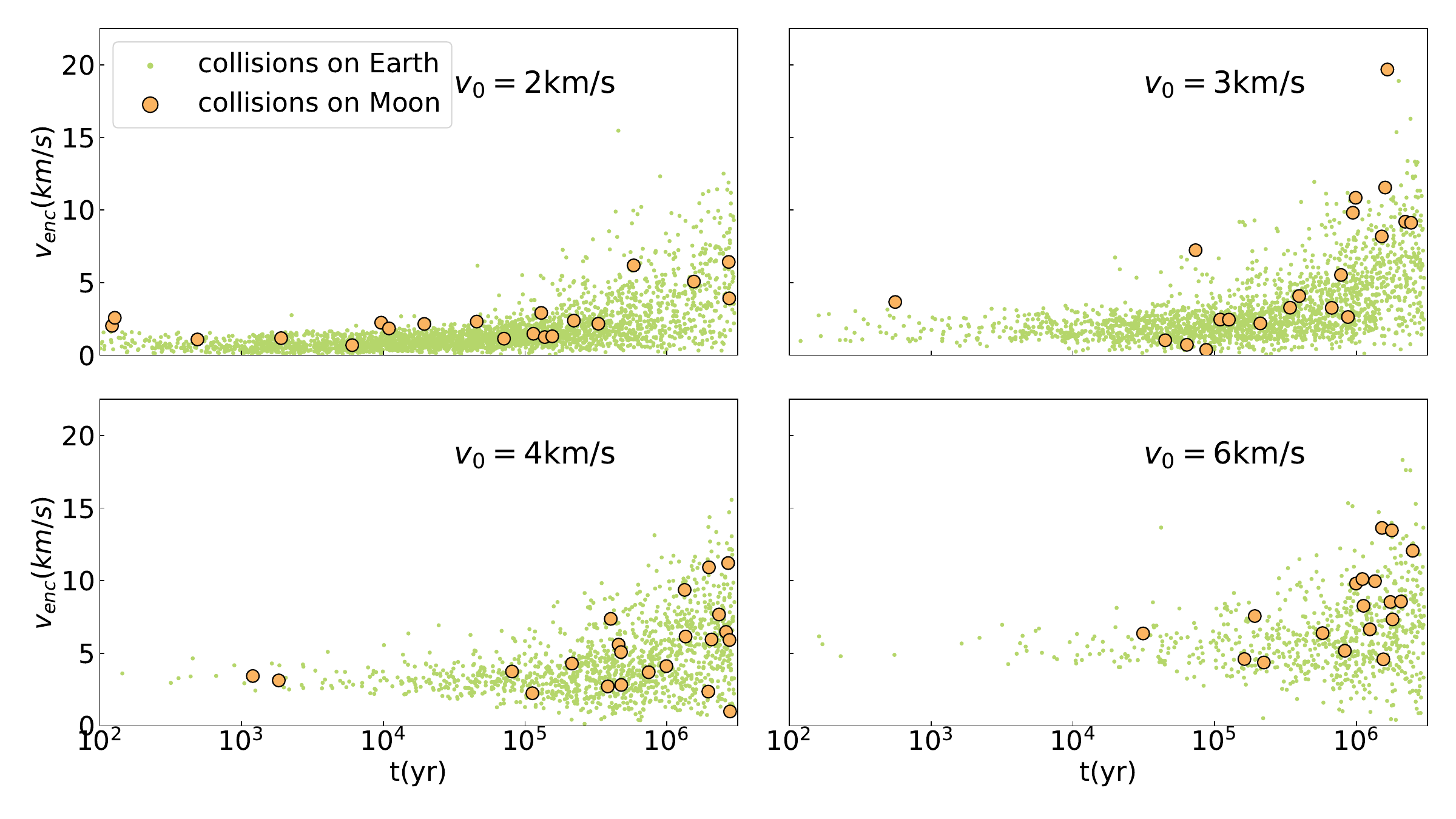}}
\caption{ Distribution of time and $v_{enc}$ for collision events on Earth-Moon system. The x-axis represents time on a logarithmic scale, while the y-axis shows the $v_{enc}$, which indicates the velocities of impactors as they enter the Hill radius of the Earth-Moon system. For clarity, impacts on the Moon are highlighted with larger markers.}
\label{fig:collision_vt}
\end{figure*}

As the orbits of the ejecta become progressively excited, both their flux and velocity upon returning to the Earth-Moon system are consequently affected. Our simulations recorded substantial impact events on terrestrial planets, with Earth receiving the majority (67.33\%), followed by Venus (31.15\%) and the Moon (0.83\%). Based on the recorded impact events, we calculated the $v_{enc}$ of the objects upon entering the Hill radius of the Earth-Moon system. This velocity was subsequently used to estimate the probability of impacts on the leading versus trailing hemisphere of the Moon. Figure~\ref{fig:collision_vt}  illustrates the impact time and the distribution of $v_{enc}$ for the collision events on the Earth and the Moon.

The temporal distribution of impacts reveals distinct phases of ejecta evolution. The initial phase ($<10$\,kyr) shows minimal velocity evolution, with $v_{enc}$ closely reflecting initial conditions. Low-velocity ejecta ($v_0$ = 2\,km/s) dominate this early impact flux. On the other hand, a strong correlation exists between $v_{enc}$ and $v_0$, showing that impactors returning to the Earth-Moon system within this phase partly retain kinematic memory of their launch conditions.

The timescale for the orbital excitation of ejecta is $\sim$100\,kyr, during which the $v_{enc}$ gradually increases. Beyond 1\,Myr, the $v_{enc}$ distribution becomes homogeneous across all $v_0$ groups. This velocity evolution is directly correlated with the $\Delta v$ progression shown in Fig.~\ref{fig:v_dist}, confirming the consistency between these two metrics as diagnostic indicators of impactor dynamics.

\subsection{Impact flux and the L/T}

In this study, we individually simulated scenarios across different $v_0$ values and subsequently integrated their results through fitting and weighting procedures to derive comprehensive conclusions. Here, we summarize the basic information from several simulations in Table \ref{tab:ltr_sim}. As shown in the second row of Table \ref{tab:ltr_sim}, $v_L$ is derived from $v_0$ (when the ejecta is located at twice the lunar radius). The third row presents the launch frequency, which follows a power-law distribution proportional to $v_L^{-4}$ \citepads{Jiao2024}, with normalization such that the relative frequency equals 1 when $v_L$ = 2.38\,km/s (the Moon's escape velocity). Row 4 reports the successful escape counts, referring to the surviving ejecta population at $t$ = 100\,yr.

\begin{table}[!htb]
    \caption{Proportion of ejecta at different $v_0$ and the total number of collisions.}
    \centering
    \begin{tabular}{c|cccccc}
    \hline
        $v_0$\,(km/s) & 2     & 2.5   & 3     & 4     & 5     & 6 \\ 
        $v_L$\,(km/s) & 2.614 & 3.014 & 3.440 & 4.340 & 5.276 & 6.232 \\ \hline
        Frequency    & 0.687   & 0.389 & 0.229 & 0.090 & 0.041 & 0.021 \\ 
        Escape       & 7984     & 9595  & 9971  & 9988  & 9994  & 9994  \\ \hline
                     & \multicolumn{6}{c}{Collision counts}  \\ \hline
        Earth & 3222     & 2786  & 2103  & 1319  & 981  & 749  \\ 
        Moon  & 33       & 31    & 21    & 23    & 12   & 18   \\ \hline
    \end{tabular}
    \label{tab:ltr_sim}
\end{table}

\begin{figure}[!htb]
\centering
\resizebox{\hsize}{!}{\includegraphics{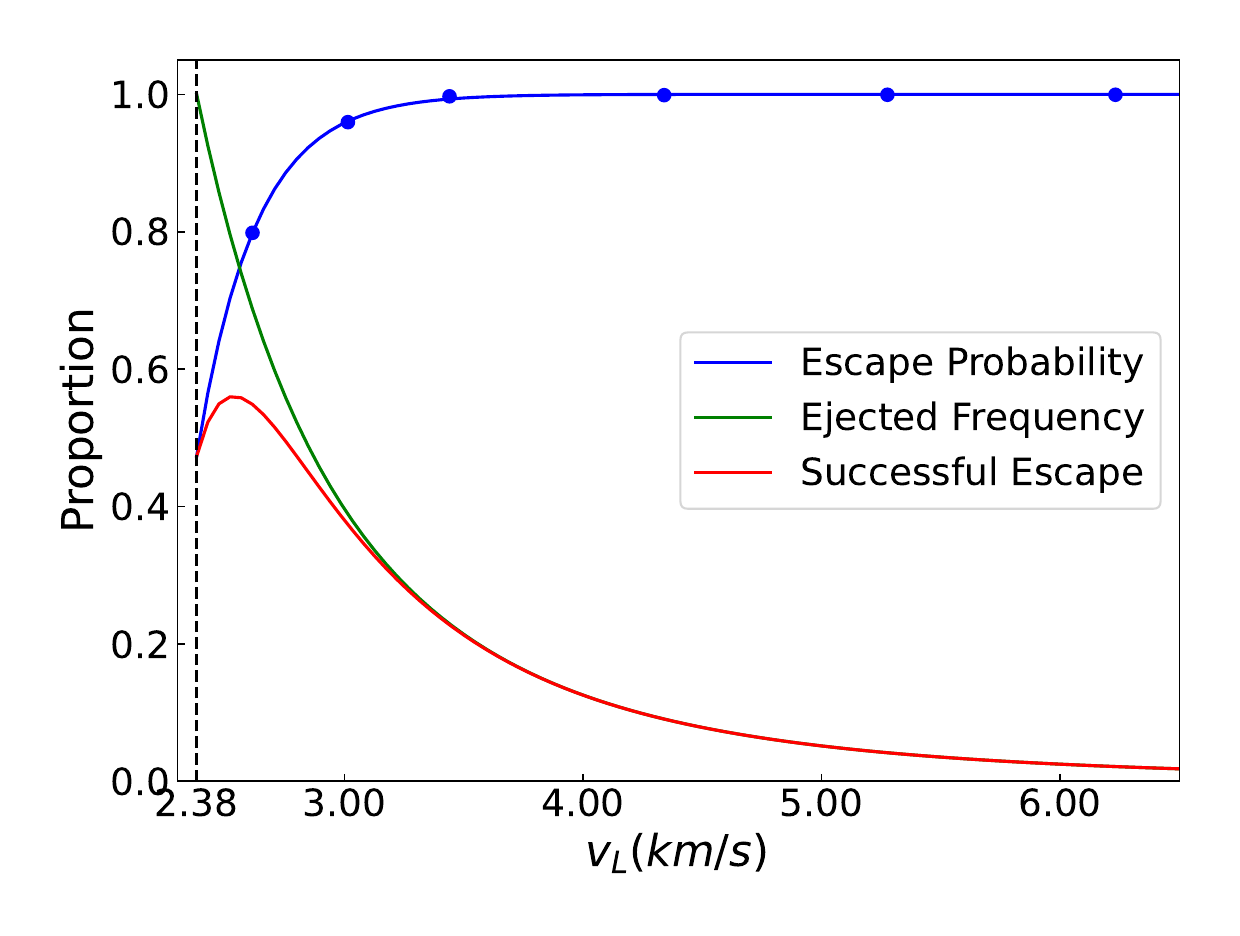}}
\caption{ Relationship between the proportion of ejecta escaping the Earth-Moon system and their $v_L$. The launch frequency is provided by \citetads{Jiao2024}, while the escape probability is derived from simulation results (represented by blue points). The dashed line on the left indicates the Moon's escape velocity.}
\label{fig:vl_dist}
\end{figure}

As our primary focus is on the ejecta that have the capacity to escape the Earth-Moon system's Hill radius, Table \ref{tab:ltr_sim} reveals that the probability of ejecta failing to escape can be well-fitted with an exponential function. Meanwhile, the probability of escaping can be expressed as $P_{escape}=1-9139.4 \exp(-4.1v_L)$, as shown by the blue line in Fig.~\ref{fig:vl_dist}. Thus, the population of successfully escaped ejecta follows the distribution $f(v_L)=45.83v_L^{-4}(1-9139.4 \exp(-4.1v_L))$, where the coefficient 45.83 ensures that the integral of this function from 2.38 to infinity equals 1. This $f(v_L)$ serves as the weighting function for all subsequent calculations of impact flux, $v_{enc}$, and L/T.

From Fig.~\ref{fig:collision_vt}, it is evident that small bodies with different $v_0$ values exhibit varying decay timescales for their impact frequencies. Slow ejecta tend to produce numerous impacts during the early phase, but their collision frequency decays rapidly, while fast ejecta behave in the opposite way. Through empirical testing, we find that the cumulative number of Earth-Moon system impacts at a given time $t$, denoted as $C(t)$, can be well described by

\begin{equation}
    C(t)= m(1 - \exp(\frac{-c  t^k}{m}))
    \label{eq:cdf}
.\end{equation}

\begin{figure}[!htb]
\centering
\resizebox{\hsize}{!}{\includegraphics{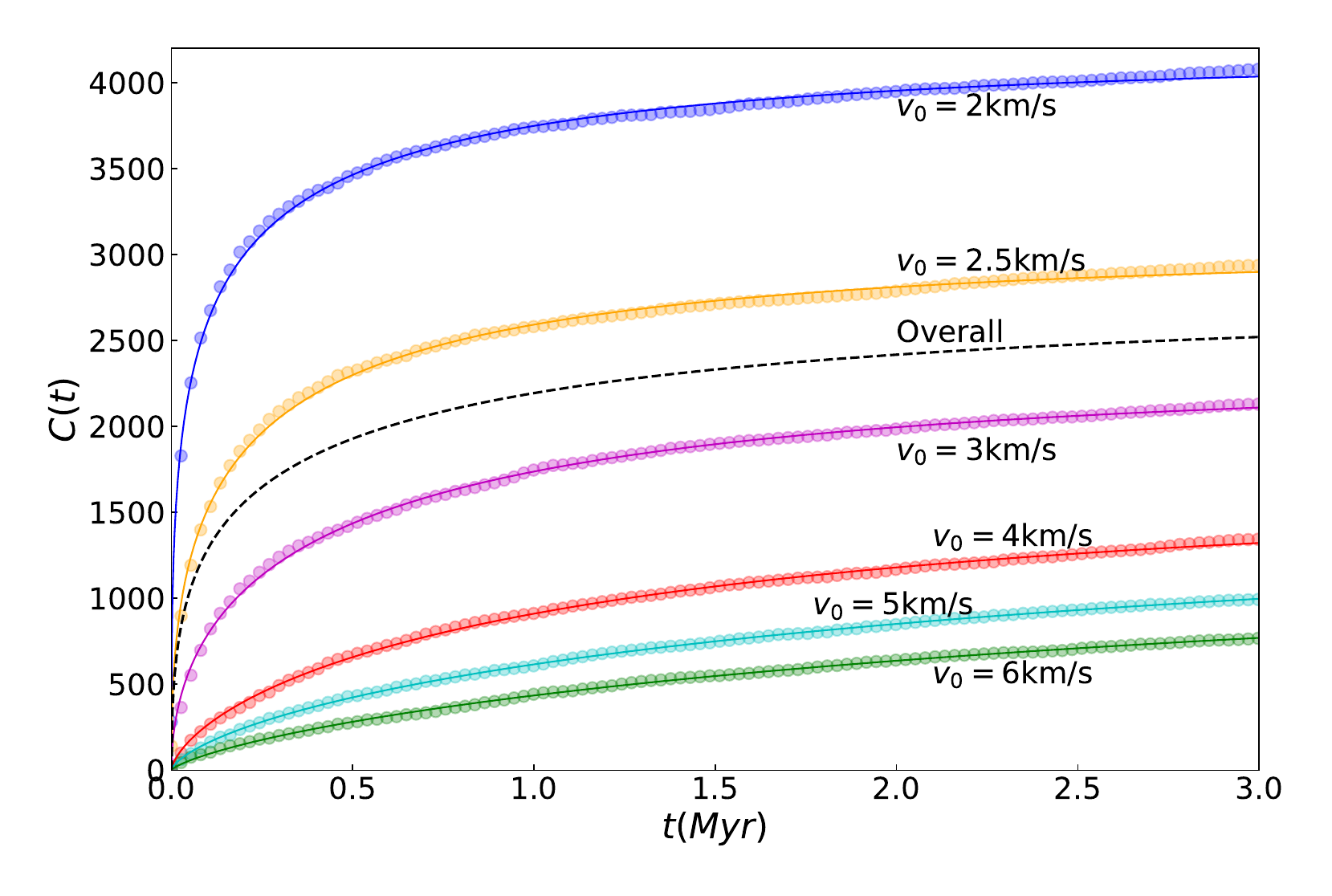}}
\caption{Cumulative impact count of lunar ejecta on the Earth-Moon system as a function of time. The light-colored points represent the simulated cumulative impact counts, while the solid lines correspond to the fitted results based on Eq. \ref{eq:cdf}. The black dashed line represents the overall result obtained after weighting different $v_0$.}
\label{fig:cdf_fit}
\end{figure}

The derivative of Eq. \ref{eq:cdf} is given by $\frac{dC(t)}{dt}= ckt^{k-1}\exp(\frac{-ct^k}{m})$, and for small $t$, the collision rate is $\frac{dC(t)}{dt} \approx ckt^{k-1}$. Here, parameter $c$ governs the baseline collision frequency, while $k$ is a value between 0 and 1 that characterizes the decay of the function. Consequently, at early simulation times, the cumulative impact count approximately follows a power law $C(t) \approx ct^{k}$, which aligns with the typical functional form \citepads[e.g.,][]{CastroCisneros2025b}. However, a simple power function does not converge as $t \rightarrow \infty$. To account for finite ejecta populations and improve late-stage fitting accuracy, an additional parameter $m$ is introduced. This modification ensures that for large $t$, $C(t)$ asymptotically approaches $m$ from below. In Fig.~\ref{fig:cdf_fit}, we illustrate the evolution of the $C(t)$ over time for different $v_0$, along with the fitting results from Eq. \ref{eq:cdf}.

Since we  previously accounted for the escape probability of ejecta, in this fitting procedure, we normalized the impact count for each $v_0$ group by dividing it by the corresponding escape probability; namely, by assuming 10,000 small bodies of each $v_0$ successfully escape the Earth-Moon system. According to Eq. \ref{eq:cdf}, the number of ejecta impacting the Earth-Moon system between $t_0$ and $t_1$ is given by

\begin{equation}
    C(t_1)-C(t_0)=\int_{2.38}^{\infty} f(v_L) \times (C(v_L,t_1)-C(v_L,t_0)) d{v_L}
    \label{eq:flux}
.\end{equation}

At this step, all parameters ($c$, $k$, and $m$) in Eq. \ref{eq:cdf} are fitted as a sigmoid function of $v_L$. Consequently, in Eq. \ref{eq:cdf}, we express the cumulative impact count as a function of both $v_L$ and $t$. In Fig.~\ref{fig:cdf_fit}, the cumulative impact count integrated over different $v_L$ is represented by the black dashed line. When calculating the L/T from $t_0$ to $t_1$, we further need to fit the L/T as a function of $v_L$. Therefore, the overall L/T during this period can be expressed as
\begin{equation}
    \frac{L(t_1)-L(t_0)}{T(t_1)-T(t_0)}=\frac{\int_{2.38}^{\infty} f(v_L) \times (C(v_L,t_1)-C(v_L,t_0)) \times \frac{L/T(v_L)}{1+L/T(v_L)} d{v_L}}{\int_{2.38}^{\infty} f(v_L) \times (C(v_L,t_1)-C(v_L,t_0)) \times \frac{1}{1+L/T(v_L)} d{v_L}}
    \label{eq:ltr}
,\end{equation}
where $L(t)$ and $T(t)$ represent the cumulative impacts received by the leading side and trailing side respectively up to the time, $t$. 

Based on the above methodology, we calculated the flux and $v_{enc}$ of lunar ejecta with different $v_0$ for collisions on the Earth-Moon system across various time intervals, as well as the resulting L/T, as shown in Table \ref{tab:ltr_cal}. Using Eqs. \ref{eq:flux} and \ref{eq:ltr}, we summarize the weighted aggregate results in the final column. We note that while the upper limits of integration in Eqs. \ref{eq:flux} and \ref{eq:ltr} are set to infinity, in practical computations, this upper bound is truncated at 10\,km/s. This cutoff ensures the inclusion of the vast majority of lunar ejecta and faster ejecta do not significantly influence the results presented in this study.

\begin{table*}[!htb]
    \caption{Impact characteristics across different $v_0$ and different time intervals.}
    \centering
    \begin{tabular}{c|cccccc|c}
    \hline
        $v_0$\,(km/s) & 2     & 2.5   & 3     & 4     & 5     & 6 & \\ 
        $v_L$\,(km/s) & 2.614 & 3.014 & 3.440 & 4.340 & 5.276 & 6.232 & \\ \hline
                         & \multicolumn{6}{c|}{Impact frequency $(kyr^{-1})$} & Collision counts    \\ \hline
        100\,yr$\sim$10\,kyr & 100.30& 52.02& 18.29& 4.75 & 2.53 & 1.92 & 577.6 \\ 
        10\,kyr$\sim$100\,kyr& 12.32 & 10.20& 6.70 & 2.28 & 1.46 & 0.87 & 706.6 \\ 
        100\,kyr$\sim$1\,Myr  & 0.98 & 1.16  & 1.06 & 0.73 & 0.50 & 0.38 & 905.3 \\ 
        1\,Myr$\sim$3\,Myr   & 0.13 & 0.17  & 0.19 & 0.22 & 0.19 & 0.16 & 322.7 \\ 
        Overall frequency/counts&1.09& 0.94 & 0.71 & 0.45 & 0.33 & 0.26 & 2512.2 \\ \hline
                             &\multicolumn{6}{c|}{$\bar{v}_{enc}$\,(km/s)}& Aggregate $\bar{v}_{enc}$\,(km/s) \\ \hline
        100\,yr$\sim$10\,kyr & 0.56 & 0.94 & 1.83 & 3.27 & 4.37 & 5.54 & 0.79 \\ 
        10\,kyr$\sim$100\,kyr& 1.00 & 1.33 & 1.94 & 3.32 & 4.50 & 5.50 & 1.35 \\ 
        100\,kyr$\sim$1\,Myr  & 2.23 & 2.49 & 3.07 & 3.92 & 4.77 & 5.88 & 2.78 \\ 
        1\,Myr$\sim$3\,Myr   & 4.52 & 5.40 & 5.57 & 6.03 & 6.21 & 6.96 & 5.48 \\ 
        Overall $\bar{v}_{enc}$&1.49& 2.18 & 3.09 & 4.48 & 5.28 & 6.29 & 2.26 \\ \hline
                             & \multicolumn{6}{c|}{L/T}                   & Aggregate L/T \\ \hline
        100\,yr$\sim$10\,kyr & 149.11 & 25.66 & 6.67 & 2.36 & 1.85 & 1.61 & 24.88 \\ 
        10\,kyr$\sim$100\,kyr& 30.30  & 12.79 & 6.09 & 2.44 & 1.86 & 1.65 & 12.32 \\ 
        100\,kyr$\sim$1\,Myr  & 5.53   & 4.46 & 3.37 & 2.38 & 1.98 & 1.68  & 4.041\\ 
        1\,Myr$\sim$3\,Myr   & 2.39   & 2.01 & 1.91 & 1.81 & 1.75 & 1.65  & 1.997\\ 
        Cumulative L/T       & 11.68  & 6.22 & 3.66 & 2.18 & 1.87 & 1.66  & 5.928\\ \hline
    \end{tabular}       
    \tablefoot{The information about the recorded impact flux, the averaged $v_{enc}$, and the L/T for various time intervals in the simulation are presented. In the rightmost column, we summarize the aggregate results after weighting the simulation outcomes across all $v_0$ values, with the impact frequency being converted into the collision counts within the corresponding time range. The results have been normalized to a population of 10,000 escaped ejecta when calculating collision frequencies and counts.}
    \label{tab:ltr_cal}
\end{table*}

The last column of Table \ref{tab:ltr_cal} reveals that approximately 25\% of ejecta will return to the Earth-Moon system by impact within 3\,Myr, with an overall $v_{enc}$ of about 2.26\,km/s and an overall L/T of about 5.9, indicating the leading hemisphere experiences nearly six times more impacts than the trailing side. Notably, about half of these collisions occur within the first 100\,kyr. The L/T shows strong time dependence, decreasing from extremely high values ($>$20) during the initial 10\,kyr to approximately 2.0 after 1\,Myr.

As shown by Table \ref{tab:ltr_sim}, roughly 1.22\% of ejecta that returned to the Earth-Moon system ultimately impacted the Moon. For every 10,000 escaped ejecta, we projected approximately 30.6 lunar impacts, distributed as 26.2 on the leading side and 4.4 on the trailing side. This pronounced asymmetry stems from the low $v_{enc}$ of re-impacting ejecta.

\section{Conclusion and discussion}

Our simulations support the idea that re-impacting lunar ejecta constitute a population of low-velocity impactors capable of generating an extreme leading-trailing asymmetry (L/T $\approx$ 5.9). This finding provides a direct solution to the long-standing discrepancy between the observed lunar crater distribution (L/T $\approx$ 1.4–1.9; e.g., \citeads{Morota2003, Oberst2012, Kawamura2011, Williams2018}) and theoretical models based solely on NEO impactors (L/T $\approx$ 1.14; e.g., \citeads{Gallant2009, Ito2010, LeFeuvre2011, Wang2016}). By considering a mixed impactor population, we find that if lunar ejecta contribute approximately 15.4\% of the total impactors, the resulting net L/T ratio aligns with the observed value of $\sim$1.4. This implies that on the leading hemisphere, ejecta re-impacts could account for nearly 23\% of craters, underscoring their disproportionate role in shaping the asymmetric surface record. 

Furthermore, the exceptionally high L/T ratio of 3.11 observed during the Imbrian-Nectarian period \citepads{Zhao2024} could be explained if ejecta did indeed comprise up to $\sim$70\% of impactors. Such a higher proportion might indicate that large impact events were relatively frequent and violent during that period, generating a substantial amount of lunar ejecta. It should be noted that the required proportion of lunar ejecta would be somewhat overestimated when we take into account the fact that the Moon's orbital semimajor axis was smaller at that time.

The potential significance of lunar ejecta as a major impactor source can benefit from indirect support coming from meteorite studies. Many meteorites found on Earth do not exhibit spectral similarities to typical NEOs or main-belt asteroids \citepads[e.g.,][]{Vernazza2008, Leon2010}. Recent research suggests that most meteorites originate from a limited number of parent bodies and collision events \citepads[e.g.,][]{Broz2024, Granvik2024, Marsset2024}. This aligns with our model, in which a single lunar impact could generate much debris that influence the Earth-Moon system's impact flux over millions of years. The stochastic nature of such events could explain temporal variations in the impactor population and the resulting cratering asymmetry.

In previous studies, when impactors were assumed to be NEOs, the estimated proportion of lunar impacts relative to the total Earth-Moon system was approximately 4.7\% \citepads{Ito2010}. In additon, theoretical studies on lunar meteoroid impacts have suggested a lunar contribution of about 3.1\% \citepads{Pokorny2019}. However, a critical finding of our study is the low probability ($\sim$1.2\%) of a returning ejecta fragment ultimately impacting the Moon. This discrepancy is not a numerical artifact, but a fundamental physical outcome dictated by gravitational focusing. The collisional cross-section of a planet scales with $b=\sqrt{R^2+\frac{2GMR}{v^2}}$, where $v$ is the velocity of impactor and $R$ and $M$ are the radius and mass of the target body. For high-velocity objects, the lunar impact probability approaches the geometric ratio of the Moon's cross-sectional area to that of the Earth-Moon system (~6.9\%). However, for low-velocity impactors, the gravitational focusing term $\frac{2GMR}{v^2}$ becomes dominant, dramatically increasing Earth's collisional cross section relative to the Moon's. Consequently, the Moon's share of impacts drops to $\sim$1.5\% at $v$ = 5\,km/s. This mechanism inherently suppresses the proportion of lunar impact for low-velocity ejecta, a factor that must be considered when estimating the absolute flux of lunar ejecta required to explain the observed cratering record.

The size distribution of ejecta fragments introduces another layer of complexity. Our results, combined with observational trends, suggest that the L/T ratio may be size-dependent. Lunar seismic data, corresponding to craters range from several meters to several tens of meters in diameter, exhibit a more pronounced leading-trailing asymmetry \citepads{Kawamura2011} compared to radial crater data, which reveal sizes of several kilometers across \citepads{Morota2003}. Similarly, cold spot data reveal that smaller cold spots show a higher degree of asymmetry \citepads{Williams2018}. Moreover, as noted above, the proportion of lunar impacts indicated by lunar meteorites \citepads{Pokorny2019} is lower than that predicted by models using NEOs impactors \citepads{Ito2010}, suggesting that smaller meteorite impacts may involve lower velocities, consequently resulting in a higher cratering asymmetry.

This size-dependent asymmetry may arise partly because small-sized bodies inherently exhibit different orbital distributions and the lunar ejecta discussed in this study likely contribute significantly to this phenomenon. However, larger ejecta fragments tend to be launched at lower velocities \citepads{Singer2020}, which could make them more likely to impact on the leading side. The intricate interplay between impactor source, velocity, and fragment size, along with their temporal evolution, warrants further investigation. On the other hand, nongravitational effects, such as the Yarkovsky effect, may play a prominent role for small-sized objects, potentially leading to outcomes such as the circularization of their orbits \citepads[e.g.,][]{Spitale2002}. This further confirms that the asymmetric cratering on the lunar surface may be strongly influenced by the size of impactor. In addition to the observational studies mentioned above, some research has also focused on smaller-scale impacts \citepads[e.g.,][]{Cremonese2013, Horanyi2015, Szalay2016, Pokorny2019, Verkercke2025}. The leading-trailing asymmetry resulting from such micro-meteoroid collisions represents another noteworthy issue for future investigation. 

This work establishes lunar ejecta as a dynamic population that effectively shapes the lunar impact environment, moving beyond their traditional view as mere geological byproducts. Our quantitative model provides a framework for integrating this component into lunar chronology models and for re-evaluating the impact history of the Earth-Moon system \citepads[e.g.,][]{Lagain2024}.

However, key uncertainties remain. The absolute production rate of ejecta that escape the Earth-Moon system is highly dependent on the magnitude and frequency of large primary impacts, which are stochastic in nature. Future works should be focused on improving the constraints on this production function across different geological epochs. Additionally, more detailed modeling of impact dynamics will allow for current models to be refined. Future missions, such as China's Tianwen-2 mission to the potentially lunar-derived asteroid Kamo'oalewa \citepads[e.g.,][]{Sharkey2021, Jiao2024}, will provide ground-truth data that can significantly constrain the properties and dynamics of material originating from the Moon.

\begin{acknowledgements}
        This work is supported by the Science and Technology Development Fund (FDCT) of Macau (grant Nos. 0034/2024/AMJ, 0008/2024/AKP, 002/2024/SKL, 0002/2025/AKP) and the National Key Research and Development Program of China (grant No.2024YFE0201000). Li-Yong Zhou thanks the support from National Natural Science Foundation of China (NSFC, Grants No.12373081 \& No.12150009) and the China Manned Space Program with grant No.CMS-CSST-2025-A16.
\end{acknowledgements}

\bibliographystyle{aa-note}
\bibliography{AsyCra}
\end{document}